\documentclass[11pt,a4paper]{article}
\input amssymb.sty
\begin{document}
\newcommand{\Si}{\Sigma}
\newcommand{\tr}{{\rm tr}}
\newcommand{\ad}{{\rm ad}}
\newcommand{\Ad}{{\rm Ad}}
\newcommand{\ti}[1]{\tilde{#1}}
\newcommand{\om}{\omega}
\newcommand{\Om}{\Omega}
\newcommand{\de}{\delta}
\newcommand{\al}{\alpha}
\newcommand{\te}{\theta}
\newcommand{\vth}{\vartheta}
\newcommand{\be}{\beta}
\newcommand{\la}{\lambda}
\newcommand{\La}{\Lambda}
\newcommand{\D}{\Delta}
\newcommand{\ve}{\varepsilon}
\newcommand{\ep}{\epsilon}
\newcommand{\vf}{\varphi}
\newcommand{\vfh}{\varphi^\hbar}
\newcommand{\vfe}{\varphi^\eta}
\newcommand{\fh}{\phi^\hbar}
\newcommand{\fe}{\phi^\eta}
\newcommand{\G}{\Gamma}
\newcommand{\ka}{\kappa}
\newcommand{\ip}{\hat{\upsilon}}
\newcommand{\Ip}{\hat{\Upsilon}}
\newcommand{\ga}{\gamma}
\newcommand{\ze}{\zeta}
\newcommand{\si}{\sigma}
\def\bfa{{\bf a}}
\def\bfb{{\bf b}}
\def\bfc{{\bf c}}
\def\bfd{{\bf d}}
\def\bfe{{\bf e}}
\def\bff{{\bf f}}
\def\bfm{{\bf m}}
\def\bfn{{\bf n}}
\def\bfp{{\bf p}}
\def\bfu{{\bf u}}
\def\bfv{{\bf v}}
\def\bft{{\bf t}}
\def\bfx{{\bf x}}
\def\bfg{{\bf g}}
\def\bfC{{\bf C}}
\def\bfS{{\bf S}}
\def\bfJ{{\bf J}}
\def\bfI{{\bf I}}
\def\bfr{{\bf r}}
\def\bfU{{\bf U}}

\def\bfal{\breve{\al}}
\def\bfbe{\breve{\be}}
\def\bfga{\breve{\ga}}
\def\bfnu{\breve{\nu}}
\def\bfsi{\breve{\sigma}}

\def\hS{{\hat{S}}}

\newcommand{\li}{\lim_{n\rightarrow \infty}}
\def\mapright#1{\smash{
\mathop{\longrightarrow}\limits^{#1}}}

\newcommand{\mat}[4]{\left(\begin{array}{cc}{#1}&{#2}\\{#3}&{#4}
\end{array}\right)}
\newcommand{\thmat}[9]{\left(
\begin{array}{ccc}{#1}&{#2}&{#3}\\{#4}&{#5}&{#6}\\
{#7}&{#8}&{#9}
\end{array}\right)}
\newcommand{\beq}[1]{\begin{equation}\label{#1}}
\newcommand{\eq}{\end{equation}}
\newcommand{\beqn}[1]{\begin{eqnarray}\label{#1}}
\newcommand{\eqn}{\end{eqnarray}}
\newcommand{\p}{\partial}
\newcommand{\di}{{\rm diag}}
\newcommand{\oh}{\frac{1}{2}}
\newcommand{\su}{{\bf su_2}}
\newcommand{\uo}{{\bf u_1}}
\newcommand{\SL}{{\rm SL}(2,{\mathbb C})}
\newcommand{\GLN}{{\rm GL}(N,{\mathbb C})}
\def\sln{{\rm sl}(N, {\mathbb C})}
\def\sl2{{\rm sl}(2, {\mathbb C})}
\def\SLN{{\rm SL}(N, {\mathbb C})}
\def\SLT{{\rm SL}(2, {\mathbb C})}
\newcommand{\gln}{{\rm gl}(N, {\mathbb C})}
\newcommand{\PSL}{{\rm PSL}_2( {\mathbb Z})}
\def\f1#1{\frac{1}{#1}}
\def\lb{\lfloor}
\def\rb{\rfloor}
\def\sn{{\rm sn}}
\def\cn{{\rm cn}}
\def\dn{{\rm dn}}
\newcommand{\rar}{\rightarrow}
\newcommand{\upar}{\uparrow}
\newcommand{\sm}{\setminus}
\newcommand{\ms}{\mapsto}
\newcommand{\bp}{\bar{\partial}}
\newcommand{\bz}{\bar{z}}
\newcommand{\bw}{\bar{w}}
\newcommand{\bA}{\bar{A}}
\newcommand{\bL}{\bar{L}}
\newcommand{\btau}{\bar{\tau}}

\newcommand{\Sh}{\hat{S}}
\newcommand{\vtb}{\theta_{2}}
\newcommand{\vtc}{\theta_{3}}
\newcommand{\vtd}{\theta_{4}}

\def\mC{{\mathbb C}}
\def\mZ{{\mathbb Z}}
\def\mR{{\mathbb R}}
\def\mN{{\mathbb N}}
\def\mP{{\mathbb P}}

\def\frak{\mathfrak}
\def\gg{{\frak g}}
\def\gJ{{\frak J}}
\def\gS{{\frak S}}
\def\gL{{\frak L}}
\def\gG{{\frak G}}
\def\gk{{\frak k}}
\def\gK{{\frak K}}
\def\gl{{\frak l}}
\def\gh{{\frak h}}
\def\gH{{\frak H}}

\newcommand{\ran}{\rangle}
\newcommand{\lan}{\langle}
\def\f1#1{\frac{1}{#1}}
\def\lb{\lfloor}
\def\rb{\rfloor}
\newcommand{\slim}[2]{\sum\limits_{#1}^{#2}}

\newcommand{\sect}[1]{\setcounter{equation}{0}\section{#1}}
\renewcommand{\theequation}{\thesection.\arabic{equation}}
\newtheorem{predl}{Proposition}[section]
\newtheorem{defi}{Definition}[section]
\newtheorem{rem}{Remark}[section]
\newtheorem{cor}{Corollary}[section]
\newtheorem{lem}{Lemma}[section]
\newtheorem{theor}{Theorem}[section]

\vspace{0.3in}

\vspace{0.3in}
\begin{flushright}
 ITP-UH-01/06\\
 ITEP-TH-04/06
\end{flushright}
\vspace{10mm}
\begin{center}
{\Large{\bf Elliptic Schlesinger system and
Painlev{\'e} VI,
}
}\\
\vspace{5mm}

Yu.Chernyakov,$\dag$ A.M.Levin,$\diamondsuit\ddag$ M.Olshanetsky,$\dag\ddag\natural$\\
 A.Zotov$\dag$\\

\vspace{3mm}

{\it
$\dag$ - Institute of Theoretical and Experimental Physics, Moscow,\\
$\diamondsuit$ - Institute of Oceanology, Moscow,\\
$\ddag$ - Max Planck Institute of Mathematics, Bonn\\
$\natural$ - Institute of Theoretical Physics, Hannover University, Hannover}

\vspace{5mm}

\texttt{Dedicated to the centenary of the publication\\
of the Painleve VI equation
 in the Comptes Rendus \\de l'Academie des Sciences de Paris\\
 by Richard Fuchs in 1905.}

\end{center}
\vspace{5mm}

\begin{abstract}
 We construct an elliptic generalization of the Schlesinger system (ESS) with
 positions of marked points on an elliptic
 curve and its modular parameter as independent variables
 (the parameters in the moduli space of the complex structure).
  ESS is a non-autonomous Hamiltonian system
 with pair-wise commuting Hamiltonians.
 The system is bihamiltonian with respect to the linear and the quadratic Poisson
 brackets. The latter are the multi-color generalization
 of the Sklyanin-Feigin-Odeskii classical algebras.
 We give the Lax form of the ESS.
 The Lax matrix defines a connection of a flat bundle of degree one over the
 elliptic curve with first order poles at the marked points.
 The ESS is the
 monodromy independence condition on the complex structure for the linear systems related to the flat bundle.
  The case of four points for a special initial data is reduced to the Painlev{\'e} VI
 equation in the form of the Zhukovsky-Volterra gyrostat, proposed in our previous paper.
\end{abstract}


\tableofcontents

\section{Introduction}
\setcounter{equation}{0}

The Schlesinger system was introduced in \cite{Sch} is a system
of first order differential equations for $n$ matrices $\bfS^j$
$(j=1,\ldots,n)$, depending on $n$ points  $x_k\in\mC\mP^1$
\beq{S0}
\p_k\bfS^j=\frac{[\bfS^k,\bfS^j]}{x_k-x_j}\,,~(k\neq j)\,,
~~\p_k=\p_{x_k}\,,
\eq
\beq{S}
\p_k\bfS^k=-\sum_{j\ne k}\frac{[\bfS^k,\bfS^j]}{x_k-x_j}\,.
\eq
This system has the Hamiltonian form with respect to the
linear (Lie-Poisson) brackets on $\sln$. The Hamiltonian
$$
H_k=\sum_{j\neq k}\frac{\lan\bfS^k\bfS^j\ran}{x_k-x_j}\,~~(\lan\,\,\ran=\tr)
$$
defines the evolution with respect to the time $x_k$.
There exists the tau-function $\exp{\cal F}$, related to the Hamiltonians \cite{JMU}
 $$
 \p_k\ln\exp{\cal F}=H_k\,.
 $$
The Schlesinger equations are the monodromy preserving conditions for the
linear system on  $\mC\mP^1$
$$
\left(\p_z+\sum_j\frac{\bfS^j}{z-x_j}\right)\Psi=0\,.
$$
For two by two matrices and four marked points the Schlesinger system
is equivalent to the Painlev{\'e} VI equation \cite{O1}. In this case the position
of three points can be fixed as $(0,1,\infty)$ while $x_4$ play the role of an independent
variable. Due to $\SLT$ gauge symmetry we leave with second order
differential equation for the matrix element $(1,2)$ of $\bfS^4$ (see, for example, \cite{KK}).

Here we replace $\mC\mP^1$ by an elliptic curve and define a similar system (the elliptic Schlesinger system (ESS)).
In this case, in addition to the coordinates of the marked points a new independent variable appears inevitably. It is the modular parameter
of the curve, and thereby we have an additional new Hamiltonian.
The similar systems in their integrable versions were considered earlier in
\cite{RS,Ne,ST}

We reproduce the main properties of the Schlesinger system. Moreover,
we rewrite the ESS in terms of quadratic Poisson brackets. They are a multi-color
generalization of the Sklyanin-Feigin-Odesski classical algebras \cite{Skl,FO}.
In conclusion, for the four point case and the matrices of
order two we derive the Painlev{\'e} VI equation in the form of
the Zhukovsky-Volterra gyrostat, proposed in our previous paper  \cite{LOZ}.
It was established there that the non-autonomous $\SL$ Zhukovsky-Volterra gyrostat
is equivalent to the elliptic form of the Painlev{\'e} VI equation \cite{P}
proposed by P.Painlev\'{e} one year later after Fuchs (see, also, \cite{Ma}).
The corresponding isomonodromy problem on an elliptic curve is discovered
only recently \cite{Z}.
This paper is a continuation of \cite{LOZ}, though it can be read independently.

\bigskip
{\small {\bf Acknowledgments.}\\
 The work was supported by the grants NSh-1999-2003.2 of the scientific
schools, RFBR-03-02-17554  and CRDF RM1-2545. The work of A.Z. was
 partially supported by the grant MK-2059.2005.2.
We are grateful for hospitality to
 the  Max Planck Institute of Mathematics, Bonn,  and the
 Institute of Theoretical Physics of the Hannover University, where the paper was
 prepared during the visit of A.L. (MPIM) and M.O. (MPIM, ITP)}


\section{Elliptic Schlesinger system}
\setcounter{equation}{0}

\subsection{Definition}
Let $\Si_\tau=\mC/(\mZ+\tau\mZ)$  be an elliptic curve,
with the modular parameter $\tau$, $(\Im m\tau>0)$ and
$$
D_n=(x_1,\ldots,x_n)\,,~x_j\neq x_k\,,~x_k\in\Si_\tau
$$
be the divisor  of non-coincident  points with the condition
\beq{d}
\sum x_j\in(\mZ+\tau\mZ)\,.
\eq

Consider the space ${\cal P}_{n,N}^{(1)}$ of
$n$ copies of the Lie coalgebra $\gg^*\sim\sln^*$,
related to the points of the divisor.
\beq{lb}
{\cal P}_{n,N}^{(1)}=\oplus_{j=1}^n\gg_j^*\,,~~
\gg_j^*=\{\bfS^j=\sum_{\al\in\ti{\mZ}^{(2)}_N} S_\al^jt^\al
\}\,,
\eq
where $t^\al$ is the basis (\ref{db}).

Introduce three operators that act from ${\cal P}_{n,N}^{(1)}$ to the
dual space $\oplus_{j=1}^n\gg_j$
\beq{phi}
\bfI_{kj}\,:\,\gg_k^*\to\gg_j\,,~~
S^k_\ga   \mapsto(I_{kj})_\ga S^k_\ga\,,~
  (I_{kj})_\ga=\varphi_{\ga}(x_j-x_k) \,,
\eq
\beq{J}
\bfJ_{jj}\,:\,\gg_j^*\to\gg_j\,,~~
S^j_\ga\mapsto    J_\ga S^j_\ga\,,~J_\ga=E_2(\bfga)\,,
\eq
\beq{j}
\bfJ_{kj}\,:\,\gg_k^*\to\gg_j\,,~~
S^k_\ga\mapsto (J_{kj})_\ga S^k_\ga\,,~(J_{kj})_\ga=f_\ga(x_j-x_k)
\eq
where $\varphi_\ga(x)$, $E_2(\bfga)$ and $f_\ga(x)$ are defined by
(\ref{AA50}) - (\ref{fzh}).

The positions of the marked points $x_j\in D_n$, satisfying (\ref{d}), and the modular
parameter $\tau$ are local coordinates in an open cell in the moduli
space ${\cal M}_{1,n}$ of elliptic curves with $n$ marked points and play the role of times.

\begin{defi}
 The elliptic Schlesinger system (ESS) is
  the consistent dynamical system on  ${\cal P}_{n,N}^{(1)}$
with independent variables from ${\cal M}_{1,n}$
\beq{1}
\p_j\bfS^k=[\bfI_{kj}(\bfS^j),\bfS^k]\,,~(k\neq j)\,,
~~\p_k=\p_{x_k}\,,
\eq
\beq{2}
\p_k\bfS^k=-\sum_{j\ne k}[\bfI_{jk}(\bfS^j),\bfS^k]\,,
\eq
\beq{3}
\p_\tau\bfS^j=
\sum_{k\neq j}\f1{2\pi\imath}[\bfS^j,\bfJ_{kj}(\bfS^k)]
+\f1{4\pi\imath}[\bfS^j,\bfJ_{jj}(\bfS^j)]\,,
\eq
where the commutators are understand as the coadjoint action of $\gg_j$
on $\gg^*_j$.
\end{defi}
The consistency of the system will be proved below.

In the basis $t^\al\,$ $(\al\in\ti{\mZ}^{(2)}_N)$
(\ref{db}) the ESS takes the form
\beq{2.1}
\p_kS_\al^j=\sum_{\ga\in\ti{\mZ}^{(2)}_N)}\bfC(\ga,\al)S_\ga^k S^j_{\al-\ga}\varphi_{\ga}(x_j-x_k)\,,~~(k\neq j)\,,
\eq
\beq{2.1a}
 \p_kS_\al^k=\sum_{\ga\in\ti{\mZ}^{(2)}_N)}\bfC(\ga,\al)
 \sum_{j\neq k}S_{\al-\ga}^j S^k_{\ga}\varphi_{\al-\ga}(x_k-x_j)\,,
\eq
\beq{2.2}
\p_\tau S^k=\f1{2\pi\imath}\sum_{\ga\in\ti{\mZ}^{(2)}_N)}\bfC(\al,\ga)
\left(
\sum_{k\neq j}
S_{\al-\ga}^kS^j_{\ga}f_\ga(x_k-x_j) +S_\ga^kS_{-\ga}^kE_2(\bfga)
\right)\,.
\eq

\begin{rem}
Equations (\ref{2.1}), (\ref{2.1a}) are consistent with
the restriction on positions of the marked points (\ref{d})
i.e. $\,\sum_{j=1}^n\p_j\bfS^k=0$.
\end{rem}
\begin{rem}
In the  rational limit
(\ref{2.1}) and (\ref{2.1a}) pass to the standard Schlesinger system
(\ref{S0}), (\ref{S}) (see (\ref{A.3a})).
\end{rem}

As in the rational case the ESS has some fundamental properties
\begin{itemize}
  \item
  The space ${\cal P}_{n,N}^{(1)}$ is Poisson with respect to the
linear Lie-Poisson brackets on $\gg^*$
\beq{lpb}
\{S_\al^j,S_\be^k\}_1=\de^{jk}\bfC(\al,\be)S_{\al+\be}
\eq
ESS is a non-autonomous Hamiltonian system with respect to the
  linear  brackets on ${\cal P}_{n,N}^{(1)}$
  \beq{2.3}
  \p_k\bfS^j=\{H_{k},\bfS^j,\}_1\,,~~\p_k=\p_{x_k}\,,~(1,\ldots,n)\,,
  \eq
  \beq{2.4}
  \p_\tau \bfS^j=\{H_{0},\bfS^j\}_1\,,
\eq
where
\beq{2.5}
H_{k}=-\sum_{j\neq k}\lan\bfI_{kj}(\bfS^k)\bfS^j)\ran=
   - \sum_{j\neq k}\sum_{\ga\in\ti{\mZ}^{(2)}_N}
S_\ga^kS_{-\ga}^j\varphi_\ga(x_j-x_k)\,,
\eq
\beq{2.6}
H_\tau=H_0=-\f1{2\pi\imath}
\left(
\sum_{k\neq j}\lan \bfS^j \bfJ_{kj}(\bfS^k)\ran+
\sum_j\lan \bfS^j \bfJ_{jj}(\bfS^j)\ran
\right)
\eq
$$
=-\f1{2\pi\imath}
\left(\sum_{k\neq j}\sum_{\ga\in\ti{\mZ}^{(2)}_N}
S_\ga^jS_{-\ga}^kf_\ga(x_k-x_j)+\sum_{ j}\sum_{\ga\in\ti{\mZ}^{(2)}_N}
S_\ga^jS_{-\ga}^jE_2(\bfga)
\right)\,.
$$
The brackets (\ref{lpb}) are degenerate. The symplectic leaves are
$n$ copies of coadjoint orbits ${\cal O}_j$ $(j=1,\ldots,n)$
of $\SLN$. Let all orbits be generic,
and $c^\mu(j)$ be corresponding  Casimir functions of order $\mu$
$(\mu=2,\ldots,N)$.
The phase space of ESS is
\beq{ps1}
{\cal R}_{n,N}^{(1)}\sim {\cal P}_{n,N}^{(1)}/\{c^\mu(j)=c^{\mu}(j)_{0}\}
\sim\prod{\cal O}_j
\eq
\beq{dps1}
\dim{\cal R}_{n,N}^{(1)}=nN(N-1)
\eq
The ESS can be considered as a system of interacting non-autonomous
$\SLN$ Euler-Arnold tops, where operators (\ref{phi}), (\ref{J}), (\ref{j}) play the role of the inverse inertia tensors. \\
  \item
  The Hamiltonians satisfy the generalized Whitham equations \cite{Kr}
  \beq{w}
\p_jH_k-\p_kH_j=0\,,~~(j,k=0,\ldots,n)\,.
\eq
In other words, the flows commute and the equations (\ref{1}), (\ref{2}) and
(\ref{3}) are consistent.
  These conditions  provide the existence of the tau-function $\exp{\cal F}$
  $$
  H_{j}=\p_j{\cal F}\,,~~H_0=\p_\tau{\cal F}\,.
$$
  \item ESS is the monodromy preserving condition for  flat rank $N$
  and degree one bundles over $\Si_\tau$ with respect to deformations of its moduli.
\end{itemize}

While the first two statements can be checked directly the last one
should be considered separately. In next subsection we prove all of them by
the symplectic reduction from trivial, though infinite Hamiltonian system.


\subsection{Derivation of ESS}

Here we derive the ESS starting with a  bundle over the elliptic curve $\Si_\tau$.
Deformations of the complex structure of $\Si_\tau$ allows us to introduce the times
and the Hamiltonians.
The ESS arises on the symplectic quotient of the space of vector bundles with respect
to the action of the $\SLN$ gauge group.

\subsubsection{Vector bundles of degree one over elliptic curves}

Let  $E_N$ be a
 degree one and rank $N$  bundle  over the elliptic curve
 $\Si_{\tau_0}\sim\mC/(\mZ+\tau_0\mZ)$
  and $Conn(E_N)=\{{\cal A}\}$ be the space of its $C^\infty$ connections. It is a symplectic space with the form
$$
\om^0=\oh\int_{\Si}\lan\de{\cal A}\wedge\de{\cal A}\ran\,.
$$

Let $(z,\bz)$ be the complex coordinates on $\Si_{\tau_0}$
$$
z=x+\tau_0y\,,~\bz=x+\bar{\tau_0}y\,,~~(0<x\,,y\leq 1)\,.
$$
 For generic degree one bundles  the transition matrices corresponding to
  the two basic cycles can be chosen as
\beq{2.8a}
\begin{array}{c}
 {\cal A}(z+1,\bz+1)=Q{\cal A}(z,\bz)Q^{-1}\,,  \\
 {\cal A}(z+\tau_0,\bz+\bar{\tau}_0)=
\tilde\La {\cal A}(z,\bz)\tilde\La^{-1}+\frac{2\pi \imath}N dz\,,
\end{array}
\eq
where $\tilde\Lambda(z,\tau)=-\bfe_N\bigl(-z-\frac{\tau_0}2\bigr)\Lambda$ and
 $Q,\La$ (\ref{q}), (\ref{la}).
It means that there are no moduli parameters for  degree one bundles.

The complex structure on $\Si_{\tau}$ allows us to introduce the complex structure on  $Conn(E_N)$.
Let
$$
d'=\p+A\,,~~d''=\bp+\bA\,,~~~(\p=\p_z\,,\,\bp=\p_{\bz})
$$
 be the corresponding components of the connection ${\cal A}$.

In addition, we fix a quasi-parabolic structure
at $n$ marked points. It means that $ A$ has simple poles at the
marked points and
$$
Res A|_{z=x_j^0}=\bfS^j=g^{-1}\bfS^j_0g\in {\cal O}_j\subset\gg_j^*
$$
while $\bA$ is regular.
The symplectic form acquires the additional Kirillov-Kostant terms
\beq{2.7}
\om^0=\int_{\Si}\lan\de A\wedge\de \bA\ran-\sum_{j=1}^n
\lan\bfS^j_0g_j^{-1}\de g_jg_j^{-1}\wedge\de g_j\ran\,,~g_j\in\SLN\,.
\eq
We denote the set $Conn(E_N)$ with the quasi-parabolic structure
at the marked points as $\ti{\cal R}_{N,\tau,n}^{(1)}(\bfS^j_0)$.

In fact, we will work with the larger space
$$
\ti{\cal P}_{n,N}^{(1)}=\{Conn(E_N)\,\,;\oplus_{j=1}^n\gg^*_j\}=
\{(A,\bA)\,,\bfS^j\,,(j=1,\ldots,n)\}
$$
equipped with the Poisson brackets
\beq{pb}
\{A_\al,\bA_\be\}=\de_{\al,-\be}\,,
\eq
\beq{pb1}
\{S^j_\al,S^k_\be\}=\de_{jk}\bfC(\al,\be)S_{\al+\be}\,.
\eq
By fixing the values of the Casimir functions to come down to $\ti{\cal R}_{N,\tau,n}^1(\bfS^j_0)$.

\subsubsection{Introducing Hamiltonians by
deformation of complex structure}

Deform the complex structure as
\beq{2a.2}
\left\{
\begin{array}{l}
 w=z-\ep(z,\bz)\,,  \\
\bar{w}=\bz\,;
\end{array}
\right.
dw=(1-\p\ep)dz-\bp\ep d\bz\,.
\eq
The Beltrami differential
$$
\mu=\frac{\bp \ep(z,\bz)}{1-\p \ep(z,\bz)}
\left(\frac{\p}{\p z}\otimes d\bz\right)
\,,~~(\bp=\p_{\bz})
$$
defines the new holomorphic structure -
the deformed antiholomorphic operator annihilates $dw$, while
the antiholomorphic structure is kept unchanged
$$
\p_{\bar{w}}=\bp+\mu\p,~~\p_w=\p.
$$
In addition, assume that $\mu$ vanishes
at the marked points $\mu(z,\bz)|_{x^0_j}=0$.

We specify the dependence of $\mu$ on the positions of
the marked points in the following  way.
Let ${\cal U}'_j\supset{\cal U}_j$ be two vicinities
 of the marked point $x_a$
such that ${\cal U}'_j\cap{\cal U}'_k=\emptyset$ for $j\neq k$.
Let $\chi_j(z,\bz)$ be a smooth function
$$
\chi_j(z,\bz)=\left\{
\begin{array}{cl}
1,&\mbox{$z\in{\cal U}_j$ }\\
0,&\mbox{$z\in\Si_g\setminus {\cal U}'_j.$}
\end{array}
\right.
$$
Introduce times related to the positions of the
marked points $t_j=x_j-x_j^0$. Then
\beq{17.2}
\mu_j=t_j\mu_j^0=t_j\bp\chi _j(z,\bz)\,,~~t_j=x_j-x_j^0\,.
\eq
The dependence of the modular parameter takes the form
\beq{2a.6}
\mu_\tau=t_\tau\mu_0^0=\frac{t_\tau}{\tau_0-\bar{\tau}_0}
\bp(\bz-z)
(1-\sum_{j=1}^n\chi_j(z,\bz))\,,
~~t_\tau=\tau-\tau_0\,.
\eq
The functions $\mu_j^0$ $\,(j=0,\ldots,n)$ can be considered as a basis in a big cell ${\cal M}^0_{1,n}$ of the moduli space ${\cal M}_{1,n}$.
The introduced above times play the role of coordinates in this
basis
\beq{me}
\mu=t_\tau\mu_\tau^0+\sum_{j=1}^nt_j\mu_j^0\,.
\eq

We deform $\om^0$ by means of the Beltrami differentials in a such way that it  acquires  nontrivial Hamiltonians.
Let us go to a new pair of the connection components
$$
(A,\bA)\to(A,\bA'=\bA-\mu A)
$$
It changes the form of $\om^0$  (\ref{2.7}) as
\beq{2.8}
\om=\om_0-
\oh\int_{\Si_\tau}\de\lan A^2\ran\de \mu\,.
\eq
Expanding $\mu$ in the basis (\ref{me}) we obtain
\beq{fo}
\om=\om^0-\sum_{j=0}^n\de\ti{H}_j\de t_j\,,~~t_0= t_{\tau}\,,
\eq
where
\beq{hj}
\ti{H}_j=\oh\int_{\Si_\tau}\lan A^2\ran\bp\chi _j(z,\bz)\,,~~(j=1,\ldots,n)
\eq
\beq{ho}
\ti{H}_0=\oh\int_{\Si_\tau}\lan A^2\ran\bp(\bz-z)
(1-\sum_{j=1}^n\chi_j(z,\bz))\,.
\eq
The form $\om$ is defined on
${\cal R}_N^1(\Si_\tau\backslash D_n)\times{\cal M}^0_{1,n}$.
The brackets (\ref{pb}), (\ref{pb1}) and the Hamiltonians $\ti{H}_j$
lead to the equations of motion
\beq{2.9}
1.\,\p_j\bA=A\mu_j^0\,,~~2.\,\p_jA=0\,,~~3.\,\p_jg_k=0\,,~~(\p_j=\p_{t_j})\,.
\eq
Evidently, these flows pairwise commute.
Moreover, we have from (\ref{pb}), (\ref{hj}), and (\ref{ho})
\beq{2.10}
\{\ti{H}_j,\ti{H}_k\}=0
\eq
\begin{rem}
It easy to see that for general non-autonomous
multi-time Hamiltonian systems,
as, for example, ESS, the commutativity of flows
amounts to the quasi-classical flatness
$$
\p_jH_k-\p_kH_j+\{H_k,H_j\}=0\,.
$$
If,  moreover, (\ref{2.10}) holds, then these conditions provide the
 existence of the tau-function\\
$\p_i\exp{\cal F} =H_i$. In particular,  the tau-function exists for the flows
(\ref{2.9}).
\end{rem}

\subsubsection{ESS as symplectic quotient}

Let ${\cal G}=\{f(w,\bw)\}$ be the group of smooth maps of $\Si_\tau$ to $\SLN$ with the quasi-periodicity
\beq{qpc}
f(w+1,\bw+1)=Q^{-1}f(w,\bw)Q\,,~~
f(w+\tau,\bw+\bar{\tau})=\ti{\La}^{-1}(w)f(w,\bw)\ti{\La}(w)\,.
\eq
Define its action on the fields as
\beq{gt}
 A \to f^{-1}\p_w f+f^{-1} Af\,,~~
 \bA\to f^{-1}\p_{\bw} f+f^{-1} \bA f\,,
\eq
$$
 g_j\to g_jf_j\,,~~f_j=f(z,\bar{z})|_{z=x_j}\,.
$$

The form $\om$ is invariant with respect to this action.
Therefore we can pass to the symplectic quotient
$$
{\cal R}_{N,\tau,n}^{(1)}(\bfS^j_0)=
\ti{\cal R}_{N,\tau,n}^{(1)}(\bfS^j_0)//{\cal G}\,.
$$
\begin{predl}
\begin{itemize}
\item The symplectic quotient is the product of the coadjoint orbits
$$
{\cal R}_{N,\tau,n}^{(1)}(\bfS^j_0)\sim \times_{j=1}^n {\cal O}_j
$$
\item
The ESS is a result of the symplectic reduction
of the system   (\ref{2.9}).
Its Hamiltonians (\ref{2.5}), (\ref{2.6}) are reduction of (\ref{hj}),
(\ref{ho}) to ${\cal R}_{N,\tau,n}^{(1)}(\bfS^j_0)$.
\item
There exists the tau-function $\exp{\cal F}$ for the ESS
$$
\p_j\exp{\cal F}=H_j\,.
$$
\end{itemize}
\end{predl}
{\it Proof}.\\
The  symplectic quotient is characterized by the conditions:\\
{\bf i.} the moment constraints
\beq{mc}
F(A,\bA)=\sum_{j=1}^n\bfS_j\de(w-x_j,\bw-\bar{x}_j)-N\de(w,\bw)t^0\,,
~~\bfS^j=g_j^{-1}\bfS_0^jg_j\,,
\eq
where $F(A,\bA)=\bp A+\p(\mu A)+[\bA,A]$.
Note that the last term in the r.h.s. of (\ref{mc}) comes from
(\ref{qpc}) and (\ref{gt}).\\
{\bf ii.} the gauge fixing
\beq{gf}
 A_{\bw}=0\,.
 \eq
 It means that any $A_{\bw}$ can be represented as the pure gauge
 $ A_{\bw}=f^{-1}[A_{\bw}]\p_{\bw} f[A_{\bw}]$.
As a result ${\cal R}_{N,\tau,n}^{(1)}(\bfS^j_0)$ is described by
the Lax matrix
$$
L= -\p_wff^{-1}+fAf^{-1}\,,~~f=f[A_{\bw}]\,.
$$
 The Lax matrix is a solution of the equation
$$
\p_{\bw} L=\sum_{j=1}^n\bfS^j\de(w-x_j,\bw-\bar{x}_j)-N\de(w,\bw)t^0
$$
with the quasi-periodicity (\ref{2.8a}). From (\ref{A.12}) and (\ref{qp}) we get
\beq{lm}
L(w)=-\f1{N} E_1(w)T_0+
\sum_{j=1}^n\sum_{\ga\in\ti{\mZ}^{(2)}_N}S_\ga^j\varphi_\ga(w-x_j)T_\ga\,.
\eq
Here, for convenience we have used the basis $T_\ga$ instead of $t^\ga$.
We stay only with finite degrees of freedom described by
the ESS variables $\bfS^j$. Thereby, the symplectic quotient ${\cal R}_{N,\tau,n}^{(1)}(\bfS^j_0)$ coincides with the phase space of the ESS
(\ref{ps1}).

The following Lemma complete the essential part of the proof.
\begin{lem}
\begin{itemize}
             \item
 The equations of motion (\ref{2.9}) on
the reduced space ${\cal R}_{N,\tau,n}^{(1)}(\bfS^j_0)$ take the Lax form
\beq{2.11}
\p_kL-\p_wM^k+[M^k,L]=0\,,~~(k=0,\ldots,n),
\eq
where
\beq{2.12}
M^k=-\sum_{\ga\in\ti{\mZ}^{(2)}_N} S_\ga^k\varphi_\ga(w-x_k)T_\ga\,,~~(k\neq 0)\,,
\eq
\beq{2.13}
M^0=-\f1{N} \p_\tau\ln\vartheta(w|\tau)T_0+\f1{2\pi\imath}
\sum_{l=1}^n\sum_{\ga\in\ti{\mZ}^{(2)}_N}S_\ga^lf_\ga(w-x_l)T_\ga\,.
\eq
             \item
 (\ref{2.11}) coincides with the ESS
(\ref{2.1}), (\ref{2.1a}), (\ref{2.2}).
           \end{itemize}
\end{lem}
{\it Proof}.\\
Substituting in the equation of motion for $A$ (\,\ref{2.9}\,(2))
$$
A=f^{-1}\p f+f^{-1} Lf
$$
and defining $M^k=-\p_kff^{-1}$ we come to (\ref{2.11}).
It follows from  (\ref{2.9}\,(1)) that $M^k$ satisfies the equation
$\p_{\bw}M^k=-L\mu^0_k$
with the same quasi-periodicity as $L$ for $j\neq 0$. To define $M^j$ we have
used (\ref{qp}) and (\ref{qpf}).
The Lax equation with $M^j$ $(j\neq 0)$ leads directly to (\ref{2.1}).
The Lax equation with $M^0$ follows from the heat equation (\ref{A.4b})
and the Calogero equation (\ref{ad2}). $\Box$

After the reduction the Poisson space $\ti{\cal P}_{n,N}^{(1)}$ passes
to ${\cal P}_{n,N}^{(1)}$ with the brackets
(\ref{pb1}). It follows from (\ref{hj}), (\ref{ho}) that the Hamiltonians $H_j$
on ${\cal P}_{n,N}^{(1)}$
can be read off from the expansion
of $\tr(L^2)$ on the basis of the elliptic functions
$$
\oh\tr(L(w))^2=\sum_{j=1}^n\left(
H_{2,j}E_2(w-x_j)+H_{1,j}E_1(w-x_j)\right)+H'_0\,,
$$
where $H_0=-\frac{1}{2\pi\imath} (H'_0-\frac{4\eta_1}{N})$
 and $\sum_jH_{1,j}=0$.
Here $H_{2,j}=\oh\sum_\ga S_\ga^jS_{-\ga}^j$ are the quadratic Casimir functions
corresponding to the orbits ${\cal O}_j$. It can be find
 that $H_{1,j}$ coincide with (\ref{2.5}), and $H_0$ with (\ref{2.6}).
 The Hamiltonians commute since their pre-images
 commute on $\ti{\cal P}_{n,N}^{(1)}$.
 Therefore, we have proved the consistency of ESS and the existence
 of the tau-function. $\Box$

 \subsubsection{Isomonodromy problem}

 Let $\Psi\in\Gamma$ be a section of a degree one vector bundle over
 $\Si_\tau$. Consider the linear system
 \beqn{2.14}
 \left\{
 \begin{array}{l}
 (\p_w+A)\Psi=0\,,\\
 (\p_{\bw}+\bA)\Psi=0\,,\\
 \p_k\Psi=0\,,~~(k=0,\ldots,n)\,.
 \end{array}
 \right.
 \eqn
The compatibility conditions of the first two equations is the
flatness condition of the bundle. The equations of motion (\ref{2.9}) are the
compatibility conditions of the last equations with the two first equations.
Let $\gamma$ be a closed path on $\Si_\tau$, $\,\Psi_\ga$ is the corresponding
transformed solution and $\Theta_\ga$ is the monodromy matrix
$$
\Psi_{\gamma}=\Psi\Theta_\ga\,.
$$
Then the last equations implies the independence of $\Theta_\ga$
on the moduli times $t_k$. Therefore, the equations of motion are the
monodromy preserving conditions.

Let $f$ be the gauge transformations $\Psi\to f\Psi$ that "kills" $A_{\bw}$.
Then (\ref{2.14}) takes the form
\beqn{2.14a}
 \left\{
 \begin{array}{l}
 (\p_w+L)\Psi=0\,,\\
 \p_{\bw}\Psi=0\,,\\
 (\p_k+M^k)\Psi=0\,,~~(k=0,\ldots,n)\,,
 \end{array}
 \right.
 \eqn
 where $L$ (\ref{lm}) and $M^k$ (\ref{2.12}),(\ref{2.13}).
The compatibility conditions of the last equations with the first one
is the ESS in the Lax form (\ref{2.11}).
They are the monodromy preserving conditions for the linear system of the
first two equations.


\section{Bihamiltonian structure of ESS}
\setcounter{equation}{0}

\subsection{Quadratic Poisson algebra}

Consider a complex space of dimension $nN^2$. We organize it in the
following way.
Attribute to the marked points of the divisor $D_n$ $n$ copies of the
 $\GLN$-valued elements
$$
x_j\to S_0^jT_0+\bfS^j=\sum_{a\in\mZ^{(2)}_N} S_a^jT_a\,.
$$
Add to this set a variable $S_0\in\mC$ and
define
$$
{\cal P}^{(2)}_{n,N}=\{S_0\,,\,(S_0^j\,,
\bfS^j\,,j=1,\ldots,n)\,|\,\sum_{j=1}^nS^j_0=0\}\,.
$$

\begin{predl}
The space  ${\cal P}^{(2)}_{n,N}$ is Poisson with respect to the quadratic  brackets
\beq{3.3}
\{S_0,S_0^j\}_2=\{S_0^j,S_0^k\}_2=\{S_\al^j,S_\al^k\}_2=0\,,
\eq
\beq{3.4}
\{S_0,S_\al^k\}_2=
\sum_{\ga\neq\al}\bfC(\al,\ga)\left(
S^k_{\al-\ga}S^k_\ga E_2(\bfga)
-\sum_{j\neq k}S^j_{-\ga}S^k_{\al+\ga}f_\ga(x_k-x_j)
\right)\,,
\eq
\beq{3.5}
\{S^k_\al,S^k_\be\}_2=\bfC(\al,\be)S_0S^k_{\al+\be}
+
\sum_{\ga\neq\al,-\be}\bfC(\ga,\al-\be)S^k_{\al-\ga}S^k_{\be+\ga}\bff_{\al,\be,\ga}
\eq
$$
+\bfC(\al,\be)S^k_0S^k_{\al+\be}(E_1(\bfal+\bfbe)-E_1(\bfal)-E_1(\bfbe))
$$
$$
-\bfC(\al,\be)\sum_{j\neq k}
[S_0^kS_{\al+\be}^j\vf_{\al+\be}(x_k-x_j)-
S_0^jS_{\al+\be}^kE_1(x_k-x_j)]\}
$$
$$
-2\sum_{j\neq k}\bfC(\ga,\al-\be)S^k_{\al-\ga}S^k_{\be+\ga}\vf_{\be+\ga}(x_k-x_j)\}
\,,
$$
where $\bff_{\al,\be,\ga}$ is defined by (\ref{fzh}). For $j\neq k$
\beq{3.6}
\{S_\al^j,S_\be^k\}_2=\sum_{\ga\neq\al,-\be}
\bfC(\ga,\al-\be)S^j_{\al-\ga}S^k_{\be+\ga}\vf_\ga(x_j-x_k)
\eq
$$
-\bfC(\al,\be)
\left(
S_0^jS_{\al+\be}^k\vf_\al(x_j-x_k)-
S_0^kS_{\al+\be}^j\vf_{-\be}(x_k-x_j)
\right)\,,
$$
and
\beq{3.7}
\{S_0^j,S_\be^k\}_2=
\left\{
\begin{array}{cc}
2\sum_\ga \bfC(\ga,-\be)S^j_{-\ga}S^k_{\be+\ga}
\vf_{\ga}(x_k-x_j)\,, & j\neq k \,,\\
-2\sum_{m\neq k}\sum_\ga \bfC(\ga,-\be)S^k_{-\ga}S^m_{\be+\ga}
\vf_{\be+\ga}(x_k-x_m)
\,, & j=k\,.
\end{array}
\right.
\eq
\end{predl}

The brackets are extracted from the classical exchange algebra
$$
\{L^{group}_1(z),L^{group}_2(w)\}_2=[r(z-w),L^{group}_1(z)\otimes L^{group}_2(w)]\,,
$$
where $r$ is the classical Belavin-Drinfeld r-matrix
$r(z)=\sum_{\ga}\vf_\ga(z) T_\ga\otimes T_{-\ga}$ \cite{BD},
and $L^{group}$ is the modified Lax operator
$$
L^{group}=\left(S_0+\sum_{j=1}^nS^j_0E_1(z-x_j)
\right)T_0+\ti{L}_j\,,~~
\ti{L}_j=\sum_\al S^j_\al\vf_\al(z-x_j)T_\al\,.
$$
The Jacobi identity for ${\cal P}^{(2)}_{n,N}$ follows from the classical
Yang-Baxter equation for $r(z)$.
The Poisson algebra ${\cal P}^{(2)}_{n,N}$ defines  the structure
 of the Poisson-Lie group on the product of $G_j$ attached to the marked points $x_j$.
The proof of Lemma will be given in a separate publication.

\begin{rem}
For $n=1$ we come to the classical Feigin-Odesski-Sklyanin algebras
\cite{Skl,FO}
\beq{3.1}
\{S_0,S_\al\}_2=
\sum_{\ga\neq\al}\bfC(\al,\ga)S_{\al-\ga}S_\ga E_2(\bfga)
\,,
\eq
\beq{3.2}
\{S_\al,S_\be\}_2=S_0S_{\al+\be}
\bfC(\al,\be)+\sum_{\ga\neq\al,-\be}\bfC(\ga,\al-\be)S_{\al-\ga}S_{\be+\ga}
\bff(\bfal,\bfbe,\bfga)\,,
\eq
\end{rem}


\subsection{Bihamiltonian structure}

The quadratic brackets on ${\cal P}^{(2)}_{n,N}$ are degenerate.
The function
$\det L(z)$ is the generating function for the Casimir functions $C^\mu(j)$
\footnote{To distinguish them from the Casimir functions of the linear algebra we
denote them by capital letters.} (see \cite{BDOZ}).
Since it is a double periodic function it can be expanded in the basis
of elliptic functions (\ref{A.2a})
\beq{6.10}
\det L(z)=C^0+\sum_j^n
C^1(j)E_1(z-x_j)+C^2(j)E_2(z-x_j)+\ldots+C^N(j)E_N(z-x_j)\,.
\eq
In particular, for the second order matrices $N=2$
\beq{c0}
C^0=S_0^2+4\eta_1\sum_{j=1}^n(S_0^j)^2+
\sum_{\ga}\left(
\sum_{j=1}^n E_2(\bfga)S^j_\ga S^j_{\ga}+2\sum_{k\neq j}S^j_\ga S^k_{-\ga}
f_\ga(x_k-x_j)\right)\,,
\eq
\beq{c1}
C^1(j)=S_0S_j+\sum_{k\neq j}S^j_0 S^k_0E_1(x_j-x_k)+
\sum_{k\neq j}\sum_{\ga}S^j_\ga S^k_{\ga}\phi_\ga(x_j-x_k)\,,
\eq
\beq{c2}
C^2(j)=(S_0^j)^2-\sum_\ga (S^j_\ga)^2\,.
\eq
Due to the condition
\beq{7.10}
\sum_{j=1}^nC^1(j)=0\,,
\eq
the number of the independent Casimir functions is $Nn$.
 The generic
symplectic leaf
$$
{\cal R}^2_{n,N}\sim{\cal P}^{(2)}_{n,N}/\{(C^\mu(j)=C^\mu(j)_{(0)})\,,~
\mu=1,\ldots,N\,,j=1,\ldots,N\}\,.
$$
has dimension
\beq{10.1}
\dim({\cal R}^2_{n,N})=nN(N-1)\,.
\eq
It coincides with the dimension of the ESS phase space
${\cal R}_{N,\tau,n}^{(1)}(\bfS^j_0)$ defined in terms of the
linear brackets.

We can extend the linear Poisson manifold ${\cal P}_{n,N}^{(1)}$
(\ref{lb}) by adding the variables $S_0\,,S_0^j$.
 In terms of the linear brackets they are
the Casimir functions and therefore preserve the phase space
${\cal R}_{N,\tau,n}^{(1)}(\bfS^j_0)$ (\ref{ps1}).

The form of brackets (\ref{3.4}), (\ref{3.7}) and the Casimir
functions (\ref{c0}), (\ref{c1}) suggests the following statement:
\begin{predl}
In terms of the quadratic brackets the ESS takes the form
$$
\p_kS_\al^j=\oh\{S_0^k,S_\al^j\}_2\,,~~(j,k=1,\ldots,n)\,,
$$
$$
\p_\tau S_\al^j=\oh\{S_0,S_\al^j\}_2\,.
$$
We have more for the second order matrices. The Casimir functions
of the quadratic brackets serve as Hamiltonians in the representations
ESS by the linear brackets
$$
\p_kS_\al^j=\{C^1(k),S_\al^j\}_1\,,~~(j,k=1,\ldots,n)\,,
$$
$$
\p_\tau S_\al^j=\f1{2\pi\imath}\{C_0,S_\al^j\}_1\,.
$$
\end{predl}
Therefore, for $N=2$ the trajectories of the ESS lie on the intersection of
the symplectic leaves of ${\cal P}^{(2)}_{n,2}$ and ${\cal P}^{(1)}_{n,N}$.
This phenomena is a manifestation of the compatibility of
 the linear and the quadratic  Poisson brackets.
 The existence of compatible Poisson structures implies  the
 bihamiltonian structure of integrable hierarchies related to these
 brackets \cite{Mag}. We don't touch this point here.


\section{Reduction to the PVI}
\setcounter{equation}{0}

Consider the rank two case $(N=2)$ with four marked points $n=4$.
We slightly change here our notations and enumerate the marked points
as $x_j$, $j=0,1,2,3$. Replace the basis $T_\al$ with the
Pauli matrices
$$
T_{(1,0)}\to\si_3\,,~~T_{(0,1)}\to\si_1\,,~~T_{(1,1)}\to\si_2\,,
$$
and the basis index $\al=1,2,3$.
As an initial data we put the marked points on $z=0$ and the half-periods
of $\Si_\tau$
$$
x_0=0\,,~x_1=\frac{\tau}{2}=\om_2\,,~x_2=\frac{1+\tau}{2}=\om_1+\om_2\,,~
x_3=\frac{1}{2}=\om_1\,,
$$
and assume that
\beq{id}
S^j_\al=\de^j_\al\ti{\nu}_\al\,,~(j=1\,,2\,,3)\,,
\eq
while $S^0_\al=S_\al$ are arbitrary.
Since for $N=2$ $\bfga\sim -\bfga$ it is not difficult to see
that
the Hamiltonians $H_j$ $\,(j=1,2,3)$ (\ref{2.5}) vanish for this configuration,
while (\ref{2.6}) assume the form
$$
H_\tau=\sum_{\ga=1,2,3}(S_\ga)^2E_2(\bfga)+S_\ga\nu_\ga'\,,~~
\nu'_\al=-\tilde{\nu}_\al
\bfe(-\om_\al\p_\tau\om_\al)\left(\frac{\vth'(0)}{\vth(\om_\al)}\right)^2\,.
$$
Therefore, the initial data (\ref{id}) stay unchanged and
we leave with the two-dimensional phase space
${\cal R}^{(1)}\subset{\cal R}^1_{4,2}$. It is described by $\bfS=(S_1\,,S_2\,,S_3)$
with the linear $\sl2$ brackets and the Casimir function
\beq{cf}
c^2=\sum_{\ga=1,2,3} S_\ga^2\,.
\eq
The equations of motion on ${\cal R}^{(1)}$ take the form
 of the non-autonomous Zhukovsky-Volterra gyrostat
\cite{LOZ}.
\beq{ur30}
\p_\tau S_\al=2\imath\epsilon_{\al\be\ga}\left(
 S_{\be}S_\ga E_2(\bfga)+\nu_\be' S_\ga\right)\,.
\eq
Here $\vec{S}=(S_1\,,S_2\,,S_3)$ is the momentum vector,
$\vec{J}=(E_2(\om_2)\,,E_2(\om_1+\om_2)\,,
E_2(\om_1))$  is the inverse inertia vector, and $\vec{\nu}'=(\nu_1'\,,\nu_2'\,,\nu_3')$ is the gyrostat momentum.
This equation has the bihamiltonian structure based on
 the generalized Sklyanin algebra \cite{LOZ}.

It was proved in \cite{LOZ} that there exists a transformation that allows to pass from the
elliptic form of the Painlev{\'e} VI \cite{P} to the  non-autonomous Zhukovsky-Volterra gyrostat (\ref{ur30}).

The Lax matrices can be read off from their representations for the ESS
(\ref{lm}), (\ref{2.13})
$$
L=
-\frac{1}{2}\p_w\ln\vth(w;\tau)\si_0+
\sum_\al (S_\al\vf_\al(w)+\nu_\al\vf_\al(w-\om_\al))\si_\al \,.
$$
$$
M=-\frac{1}{2}\p_\tau\ln\vth(w;\tau)\si_0
+\sum\limits_\al
-S_\al\frac{\varphi_1(w)\varphi_2(w)\varphi_3(w)}{\varphi_\al(w)}\sigma_\al
+E_1(w)L'\,.
$$
where $L'=\sum_\al (S_\al\vf_\al(w)+\nu_\al\vf_\al(w-\om_\al))\si_\al$.
The former matrix define the linear problem for (\ref{ur30})
in the form (\ref{2.14a}).


\section{Appendix}
\subsection{Appendix A. Elliptic functions.}
\setcounter{equation}{0}
\def\theequation{A.\arabic{equation}}

We assume that $q=\exp 2\pi i\tau$, where $\tau$ is the modular parameter
of the elliptic curve $E_\tau$.

The basic element is the theta  function:
\beq{A.1a}
\vth(z|\tau)=q^{\frac
{1}{8}}\sum_{n\in {\bf Z}}(-1)^n\bfe(\oh n(n+1)\tau+nz)=~~
(\bfe=\exp 2\pi\imath)
\eq

\bigskip

{\it The  Eisenstein functions}
\beq{A.1}
E_1(z|\tau)=\p_z\log\vth(z|\tau), ~~E_1(z|\tau)\sim\f1{z}-2\eta_1z,
\eq
where
\beq{A.6}
\eta_1(\tau)=\frac{24}{2\pi i}\frac{\eta'(\tau)}{\eta(\tau)}\,,~~
\eta(\tau)=q^{\frac{1}{24}}\prod_{n>0}(1-q^n)\,.
\eq
is the Dedekind function.
\beq{A.2}
E_2(z|\tau)=-\p_zE_1(z|\tau)=
\p_z^2\log\vth(z|\tau),
~~E_2(z|\tau)\sim\f1{z^2}+2\eta_1\,.
\eq

{\it Relation to the Weierstrass functions}
\beq{a100}
\zeta(z,\tau)=E_1(z,\tau)+2\eta_1(\tau)z\,,
~~\wp(z,\tau)=E_2(z,\tau)-2\eta_1(\tau)\,.
\eq
The highest Eisenstein functions
\beq{A.2a}
E_j(z)=\frac{(-1)^j}{(j-1)!}\p^{(j-2)}E_2(z)\,,~~(j>2)\,.
\eq

The next important function is
\beq{A.3}
\phi(u,z)=
\frac
{\vth(u+z)\vth'(0)}
{\vth(u)\vth(z)}\,.
\eq
\beq{A.300}
\phi(u,z)=\phi(z,u)\,,~~\phi(-u,-z)=-\phi(u,z)\,.
\eq
It has a pole at $z=0$ and
\beq{A.3a}
\phi(u,z)=\frac{1}{z}+E_1(u)+\frac{z}{2}(E_1^2(u)-\wp(u))+\ldots\,.
\eq
\beq{A3c}
\p_u\phi(u,z)=\phi(u,z) (E_1(u+z)-E_1(u))|_{z\to 0}=-E_2(u) \,.
\eq

{\it Heat equation}
\beq{A.4b}
\p_\tau\phi(u,w)-\f1{2\pi i}\p_u\p_w\phi(u,w)=0\,.
\eq

{\it Quasi-periodicity}

\beq{A.11}
\vth(z+1)=-\vth(z)\,,~~~\vth(z+\tau)=-q^{-\oh}e^{-2\pi iz}\vth(z)\,,
\eq
\beq{A.12}
E_1(z+1)=E_1(z)\,,~~~E_1(z+\tau)=E_1(z)-2\pi i\,,
\eq
\beq{A.13}
E_2(z+1)=E_2(z)\,,~~~E_2(z+\tau)=E_2(z)\,,
\eq
\beq{A.14}
\phi(u,z+1)=\phi(u,z)\,,~~~\phi(u,z+\tau)=e^{-2\pi \imath u}\phi(u,z)\,.
\eq
\beq{A.15}
\p_u\phi(u,z+1)=\p_u\phi(u,z)\,,~~~\p_u\phi(u,z+\tau)=e^{-2\pi \imath u}\p_u\phi(u,z)-2\pi\imath\phi(u,z)\,.
\eq

 {\it  The Fay three-section formula:}
\beq{ad3}
\phi(u_1,z_1)\phi(u_2,z_2)-\phi(u_1+u_2,z_1)\phi(u_2,z_2-z_1)-
\phi(u_1+u_2,z_2)\phi(u_1,z_1-z_2)=0\,.
\eq
Particular cases of this formula are
the  functional equations
\beq{ad2}
\phi(u,z)\p_v\phi(v,z)-\phi(v,z)\p_u\phi(u,z)=(E_2(v)-E_2(u))\phi(u+v,z)\,,
\eq
\beq{ir}
\phi(u,z_1)\phi(-u,z_2)=\phi(u,z_2-z_1)(E_1(z_1)-E_1(z_2))-\p_u\phi(u,z_2-z_1)\,.
\eq
\beq{ir1}
\phi(u,z)\phi(-u,z)=E_2(z)-E_2(u)\,.
\eq

\subsection{Appendix B.  Lie algebra $\sln$ and elliptic functions}
\setcounter{equation}{0}
\def\theequation{B.\arabic{equation}}

Introduce the notation
$$
{\bf e}_N(z)=\exp (\frac{2\pi i}{N} z)
$$
 and two matrices
\beq{q}
Q=\di({\bf e}_N(1),\ldots,{\bf e}_N(m),\ldots,1)
\eq
\beq{la}
\La=\de_{j,j+1}\,,~~(j=1,\ldots,N\,,~mod\,N)\,.
\eq
Let
\beq{B.10}
\mZ^{(2)}_N=(\mZ/N\mZ\oplus\mZ/N\mZ)\,,~~\ti{\mZ}^{(2)}_N)=
\mZ^{(2)}_N\setminus(0,0)
\eq
be the two-dimensional lattice of order $N^2$ and
$N^2-1$ correspondingly.
The matrices $Q^{a_1}\La^{a_2}$, $a=(a_1,a_2)\in\mZ^{(2)}_N$
generate a basis in the group $\GLN$, while $Q^{\al_1}\La^{\al_2}$,
$\al=(\al_1,\al_2)\in\ti{\mZ}^{(2)}_N$ generate a basis in the Lie algebra $\sln$.
More exactly, we introduce the following basis in $\GLN$.
Consider the projective representation of
$\mZ^{(2)}_N$ in $\GLN$
\beq{B.11}
a\to T_{a}=
\frac{N}{2\pi i}\bfe_N(\frac{a_1a_2}{2})Q^{a_1}\La^{a_2}\,,
\eq
\beq{AA3a}
T_aT_b=\frac{N}{2\pi i}\bfe_N(-\frac{a\times b}{2})T_{a+b}\,, ~~
(a\times b=a_1b_2-a_2b_1)
\eq
Here $\frac{N}{2\pi i}
\bfe_N(-\frac{a\times b}{2})$ is a non-trivial two-cocycle
in $H^2(\mZ^{(2)}_N,\mZ_{2N})$.
The matrices $T_\al$, $\al\in\ti{\mZ}^{(2)}_N$ generate
a basis in $\sln$.
It follows from (\ref{AA3a}) that
\beq{AA3b}
[T_{\al},T_{\be}]=\bfC(\al,\be)T_{\al+\be}\,,
\eq
where
$\bfC(\al,\be)=\frac{N}{\pi}\sin\frac{\pi}{N}(\al\times \be)$ are
 the structure constants of $\sln$.

 The Lie coalgebra $\gg^*=\sln$ has the dual basis
  \beq{db}
 \gg^*=\{\bfS=\sum_{\ti{\mZ}^{(2)}_N}S_\ga t^\ga\}\,,~~
 t^\ga=\frac{2\pi\imath}{N^2}T_{-\ga}\,,~~\lan T_\al t^\be\ran=\de_{\al}^{-\be}\,.
 \eq
It follows from (\ref{AA3b}) that $\gg^*$ is a Poisson space with the
linear brackets
\beq{A101}
\{S_\al,S_\be\}=\bfC(\al,\be)S_{\al+\be}\,.
\eq
The coadjoint action in these basises   takes the form
\beq{coad}
{\rm ad}^*_{T_\al}t^\be=\bfC(\al,\be)t^{\al+\be}\,.
\eq

Let $\bfga=\frac{\ga_1+\ga_2\tau}{N}$. Then
introduce the following  constants on $\ti{\mZ}^{(2)}$:
\beq{AA50}
\vth(\bfga)=\vth\bigl(\frac{\ga_1+\ga_2\tau}{N}\bigr)\,,
~~
E_1(\bfga)=E_1\bigl(\frac{\ga_1+\ga_2\tau}{N}\bigr)\,,
~~E_2(\bfga)=E_2\bigl(\frac{\ga_1+\ga_2\tau}{N}\bigr)\,,
\eq
\beq{ph}
\phi_\ga(z)=\phi(\bfga,z)\,,
\eq
\beq{vf}
\vf_\ga(z)=\bfe_N(\ga_2z)\phi_\ga(z)\,,
\eq
\beq{f}
f_\ga(z)=\bfe_N(\ga_2z)
\p_u\phi(u,z)|_{u=\bfga}=\vf_\ga(z)(E_1(\bfga+z)-E_1(\bfga))\,.
\eq
\beq{fzh}
\bff_{\al,\be,\ga}=E_1(\bfga)-E_1(\bfal-\bfbe-\bfga)+E_1(\bfal-\bfga)
-E_1(\bfbe-\bfga)\,.
\eq

It follows from (\ref{A.3}) that
\beq{qp}
\vf_\ga(z+1)=\bfe_N(\ga_2)\vf_\ga(z)\,,~~
\vf_\ga(z+\tau)=\bfe_N(-\ga_1)\vf_\ga(z)\,.
\eq
\beq{qpf}
f_\ga(z+1)=\bfe_N(\ga_2)f_\ga(z)\,,~~
f_\ga(z+\tau)=\bfe_N(-\ga_1)f_\ga(z)-2\pi\imath\vf_\ga(z)\,.
\eq

\small{

\end{document}